\documentclass{article} 
\usepackage{iclr2022_conference,times}
\usepackage[utf8]{inputenc}
\usepackage{amsmath,amssymb,amsfonts}
\usepackage{algorithmic}
\usepackage{graphicx}
\usepackage{textcomp}
\usepackage{wrapfig}
\usepackage{tabularx}

\usepackage{amsmath,amsfonts,bm}

\def\eqref#1{equation~\ref{#1}}

\def\1{\bm{1}}

\DeclareMathAlphabet{\mathsfit}{\encodingdefault}{\sfdefault}{m}{sl}
\SetMathAlphabet{\mathsfit}{bold}{\encodingdefault}{\sfdefault}{bx}{n}

\usepackage{hyperref}
\usepackage{url}

\title{Unleashing GPT on the Metaverse: \\Savior or Destroyer?}

\author{Pengyuan Zhou, \\
pengyuan.zhou@ece.au.dk\\
Aarhus University, Denmark}

\iclrfinalcopy
\begin{document}

\maketitle

\begin{abstract}
Incorporating artificial intelligence (AI) technology, particularly large language models (LLMs), is becoming increasingly vital for developing immersive and interactive metaverse experiences. GPT, a representative LLM developed by OpenAI, is leading LLM development and gaining attention for its potential in building the metaverse. The article delves into the pros and cons of utilizing GPT for metaverse-based education, entertainment, personalization, and support. Dynamic and personalized experiences are possible with this technology, but there are also legitimate privacy, bias, and ethical issues to consider. This article aims to help readers understand the possible influence of GPT, according to its unique technological advantages, on the metaverse and how it may be used to effectively create a more immersive and engaging virtual environment by evaluating these opportunities and obstacles.
\end{abstract}

\section{Introduction}
The metaverse~\cite{lee2021all}, a shared digital place where users may communicate and collaborate using a variety of virtual tools, has the potential to revolutionize how we study, socialize, and pass the time. Although researchers have made great strides, making a fascinating and engaging tailored metaverse experience is still a work in progress. For instance, learning to accommodate a wide variety of user interactions using natural language processing (NLP) is a significant obstacle on the path to generating a unique metaverse experience. The OpenAI-developed state-of-the-art language model GPT offers a potent answer to this situation. GPT's natural language processing features allow programmers to make a wide variety of useful applications, such as digital assistants, educational companions, unique forms of entertainment, and individualized itineraries.

While GPT shows promise as a means to tailor and enliven metaverse experiences for individual users, researchers will still need to overcome a number of obstacles. Keeping consumers interested in the metaverse requires, for example, providing a fluid and engaging user experience. For data to be used ethically, it is important to address concerns about data privacy and security.
GPT is discussed in this study to see if it can enable interesting and meaningful metaverse directions. While doing so, this work takes a look at the numerous obstacles that researchers must surmount to help readers better understand how to use GPT to create engaging and unique metaverse experiences, as well as highlight its potential pitfalls.

\section{Related Work}
Researchers have been exploring the potential of combining NLP techniques with XR in recent years. 
For instance, \cite{fagernas2021users} proposed a semi-automated NLP technique to study published reviews of different VR relaxation applications for the Oculus Go and Gear VR.
Experts in the field of education have investigated the possibility of using the metaverse to create more dynamic and engaging classroom environments~\cite{lin2015language}. \cite{uppoor2022interactive} proposed a virtual environment to provide a more engaging and rewarding experience with the help of NLP. 
\cite{yuan2022wordcraft} explores using large language models to collaborate with users to write a story. Overall, as GPT is still in its infancy, the potential/impact of GPT for the metaverse is at a very early stage waiting for further exploration.
This work sheds light on some potential future directions and discusses the challenges we need to tackle. 

\section{GPT As a Savior}



\textcolor{black}{Remarkably, the new versions of GPT (GPT-4o\footnote{\url{https://openai.com/index/hello-gpt-4o/}} and GPT-4 Turbo) provide superior capacities\footnote{\url{https://platform.openai.com/docs/overview}} to enable and accelerate metaverse development in the following aspects:
\begin{enumerate}
    \item \textbf{Multimodal understanding/generation} has the potential to push the current metaverse fast forward to its ultimate envisioned form. GPT-4o provides capabilities to understand and generate input/output including text, image, and audio, using the same neural network. With such capabilities, GPT is able to not only help users understand and analyze surroundings in the metaverse but also provide further instructions for follow-up interactions or generate feedback for users. 
    \item \textbf{API-based fine-tuning} enables personalized model-on-demand for countless metaverse services, from NLP-based virtual assistants to CV-based 3D generations. The convenience of API-based fine-tuning makes it much easier for small businesses and individual users to customize their GPT-based metaverse services tailored to personalized demands without the need to purchase and maintain expensive GPU servers. 
    \item \textbf{Function calling} allows users to describe functions to GPT models and have the model choose to output a JSON object containing expected input parameters to call one or many functions. Developers can use this capacity to extract consistent responses for API and SQL commands and call external functions, effectively accelerating and easing the development of metaverse applications. It can also facilitate users with complex interactions and real-time data reports in the metaverse. 
    \item \textbf{Code interpreter} allows GPT to write and execute programming codes to provide ready-made answers, thus saving metaverse developers considerable effort on prototyping, debugging, and data analysis.
    \item \textbf{File search} in GPT, different from keyword-based retrieval based on file metadata and indexing used in conventional search engines, enables context-aware retrieval based on content understanding. Hence, it can provide suggestive documents or text given high-level questions as well as specific answers with proactive analysis for detailed questions. Moreover, it's seamlessly integrated into the interactive chats. File search in GPT enables search-related applications such as search engines, recommendation systems, and chatbots, to provide much more insightful information while avoiding breaking the sense of presence via its seamless integration with chat functions.
    \item \textbf{Long context} enables continuous long interactions, allowing NPCs to exhibit more consistent and nuanced behaviors and interactions, or the virtual assistants to have more contextually relevant and personalized responses based on remembering the long interactions. It benefits use cases related to narrative and interaction consistency, such as gaming, training, and educational applications in the metaverse. Notably, several LLMs nowadays support similar or even longer contexts than GPT-4.
    \item \textbf{Multi-Agent} GPTs, or multi-agent LLMs in general, enabled by the communication and in-context learning capabilities of GPTs, improve the planning and reasoning ability of single GPT via discussion and debating between distinct agents specialized to diverse capabilities and roles. Its capability for complex problem-solving and world simulation is a powerful tool to run and govern large-scale metaverse platforms such as VR society or digital twin platforms. 
\end{enumerate}	
Note that as AI is rapidly advancing, more LLMs besides GPT have emerged with good performances, some even comparable to GPT in some aspects. Hence, \textit{many other LLMs will join GPT as accelerators and enablers for metaverse development in the near future.}}

Next, we outline 19 examples under 4 categories according to their purposes as summarized in Figure~\ref{fig:savior}. Each category provides illustrative cases for better understanding. We highlight the advantages specifically attributed to the capabilities mentioned above.

\subsection{Education}
\begin{wrapfigure}{r}{0.4\textwidth}
\begin{center}
    \includegraphics[width=.38\textwidth]{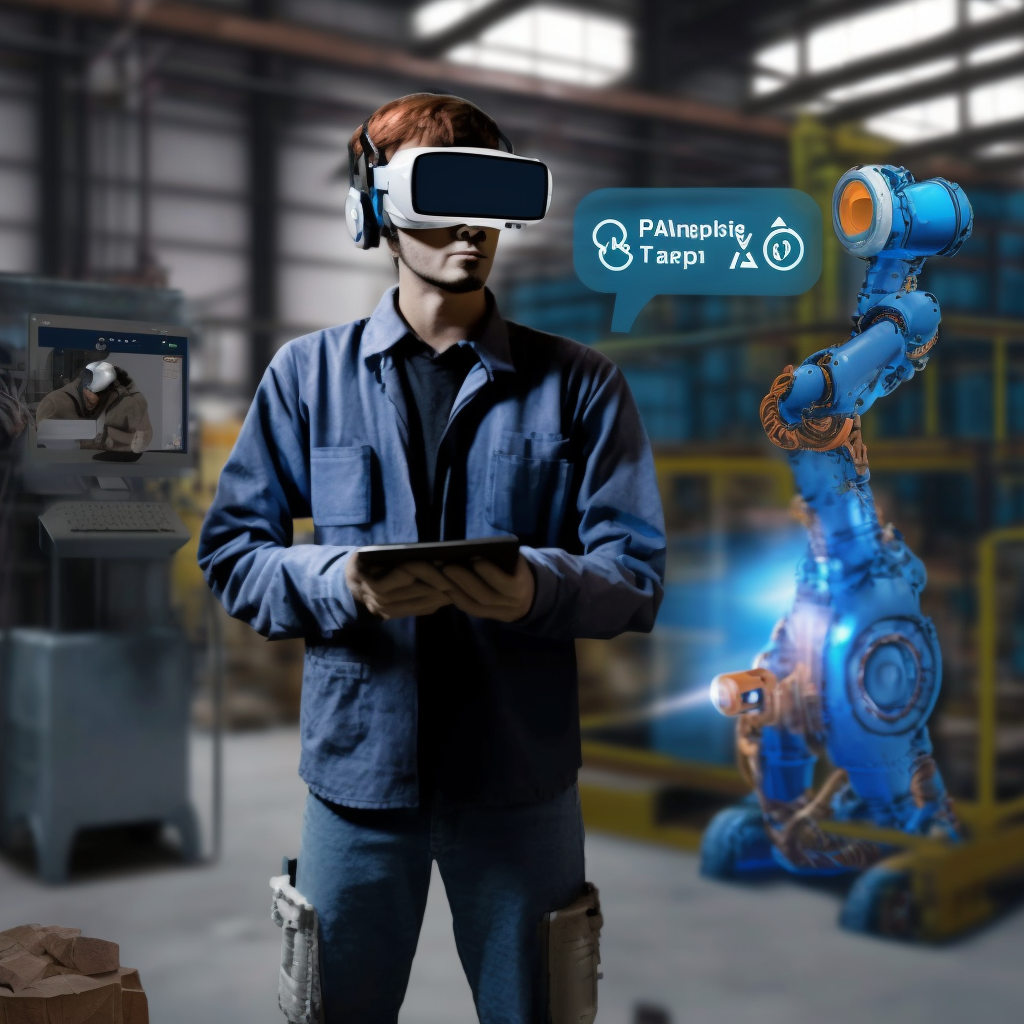}
\end{center}
    \caption{Interactive training companions.}
    \label{fig:training}
\end{wrapfigure}

\subsubsection{Interactive Educational Companions. (Figure~\ref{fig:training})}

GPT enables programmers to make engaging learning tools that may be used to teach users new knowledge and abilities. These aids can modify teaching strategies based on the individual's performance. The educational companion can, for instance, offer further explanations and examples if the user is having trouble grasping a certain concept, and it will continue to do so until the user exhibits an in-depth mastery of the material.
Students can augment their classroom learning with interactive educational companions to practice their abilities and receive instant feedback~\cite{kok2022virtual}. With the help of the companions, educators can design individualized lesson plans to address each student's unique strengths and weaknesses.
Furthermore, from math and physics to history and literature, interactive educational companions can be utilized to teach a wide variety of subjects. 
They can even be used to learn a new skill or hobby, like gardening or cooking, or in-house employee training purposes~\cite{10042870}, making the process of learning new procedures or processes in the workplace more interesting and engaging for the trainee, as illustrated in Figure~\ref{fig:companion}.

Hence, the tailored, dynamic, and engaging learning experiences made possible by GPT-created interactive educational companions have the potential to completely transform the way we learn.
\begin{figure}[!hb]
   \begin{minipage}{0.48\textwidth}
     \centering
     \includegraphics[width=.7\linewidth]{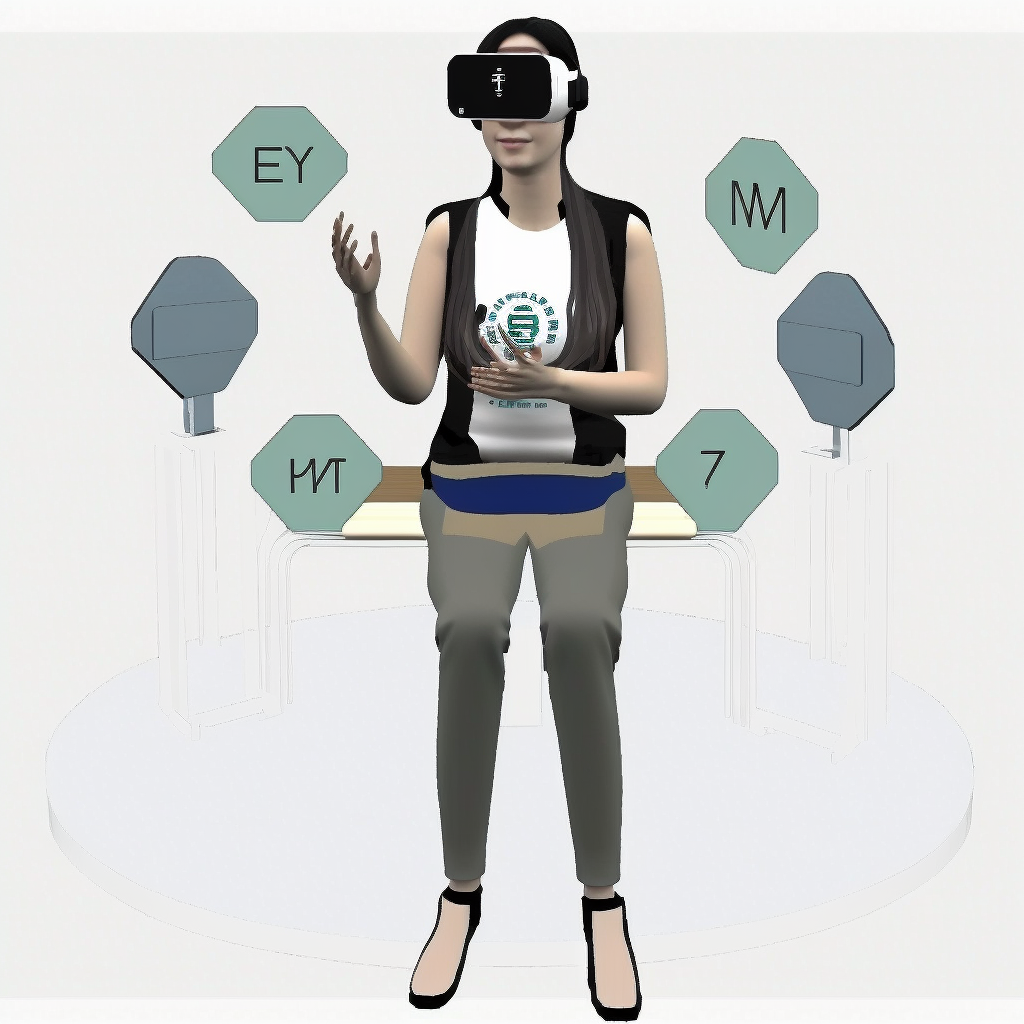}
     \caption{Interactive language companions.}\label{fig:language}
   \end{minipage}\hfill
   \begin{minipage}{0.48\textwidth}
     \centering
     \includegraphics[width=.7\linewidth]{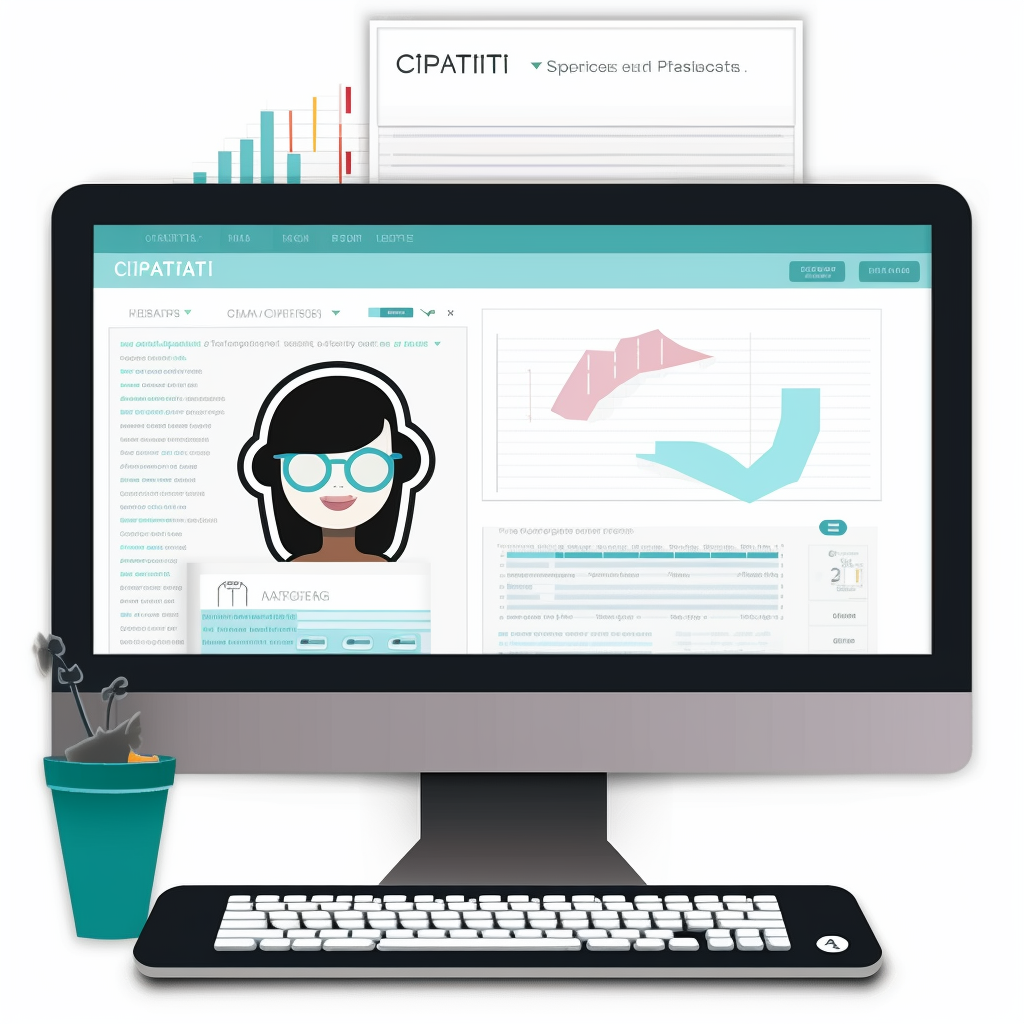}
     \caption{Virtual online class assistant.}\label{fig:online}
   \end{minipage}
\end{figure}
\subsubsection{Language Learning Companions. (Figure~\ref{fig:language})}

With GPT, researchers can make language-learning assistants that can engage with learners in the virtual world~\cite{uppoor2022interactive,parmaxi2023virtual}.
Depending on the learner's area(s) of proficiency and improvement, language learning companions can provide constructive criticism. Accurate feedback on pronunciation, grammar, and vocabulary from them is invaluable. To maximize efficiency and effectiveness, this feedback can be adjusted based on the learner's current skill level and preferred method of instruction.
Gamification features can be incorporated into language learning companions to keep students interested and motivated. Some examples of these components are tests, quizzes, and achievements. Language learning companions powered by GPT make studying a new language interesting and engaging, which in turn improves learners' persistence.

\subsubsection{Virtual assistants for online classes. (Figure~\ref{fig:online})}
Helping customers find their way through online learning platforms is just the beginning; GPT also offers individualized support in the shape of reminders, study guides, and the like. Having this kind of support can keep pupils on track and ultimately help them do better in school. GPT-enabled virtual assistants can also help instructors manage their online classrooms by responding to students' questions and delivering grades and comments in an automatic fashion. Because of this, educators may have more time to focus on guiding students.

\subsection{Entertainment}
\subsubsection{Interactive Storytelling. (Figure~\ref{fig:story})}

In virtual environments, GPT can be used to automatically produce text for use in interactive narratives~\cite{yuan2022wordcraft}. By using GPT, a user may create vivid descriptions of one's own virtual world and its inhabitants, which will enhance the narratives. Creators of massively multiplayer online role-playing games (MMORPGs) can use these features to make games that stand out from the crowd. \textcolor{black}{The multi-agent debation capability allows further narrative refinement from different players' perspectives.}

\begin{figure}[!htb]
   \begin{minipage}{0.48\textwidth}
     \centering
     \includegraphics[width=.7\linewidth]{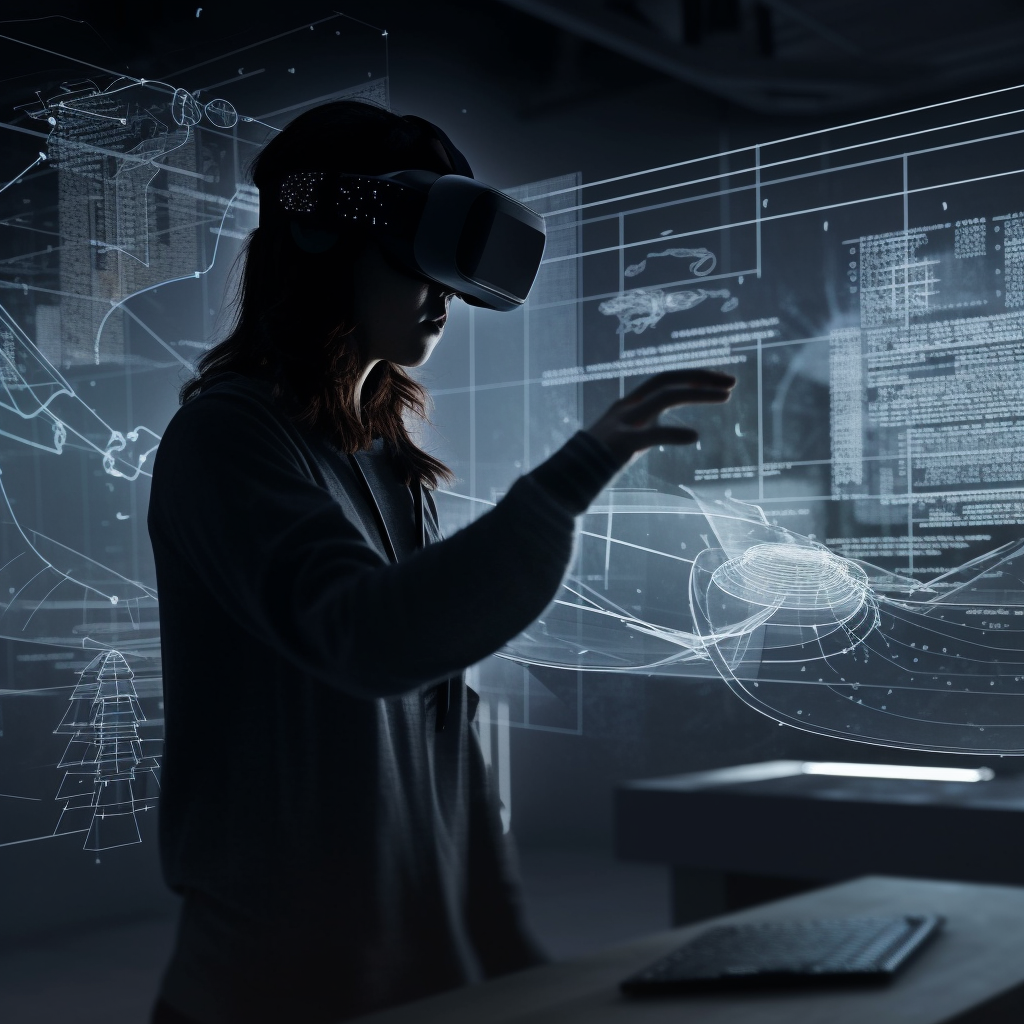}
     \caption{Interactive storytelling.}\label{fig:story}
   \end{minipage}\hfill
   \begin{minipage}{0.48\textwidth}
     \centering
     \includegraphics[width=.7\linewidth]{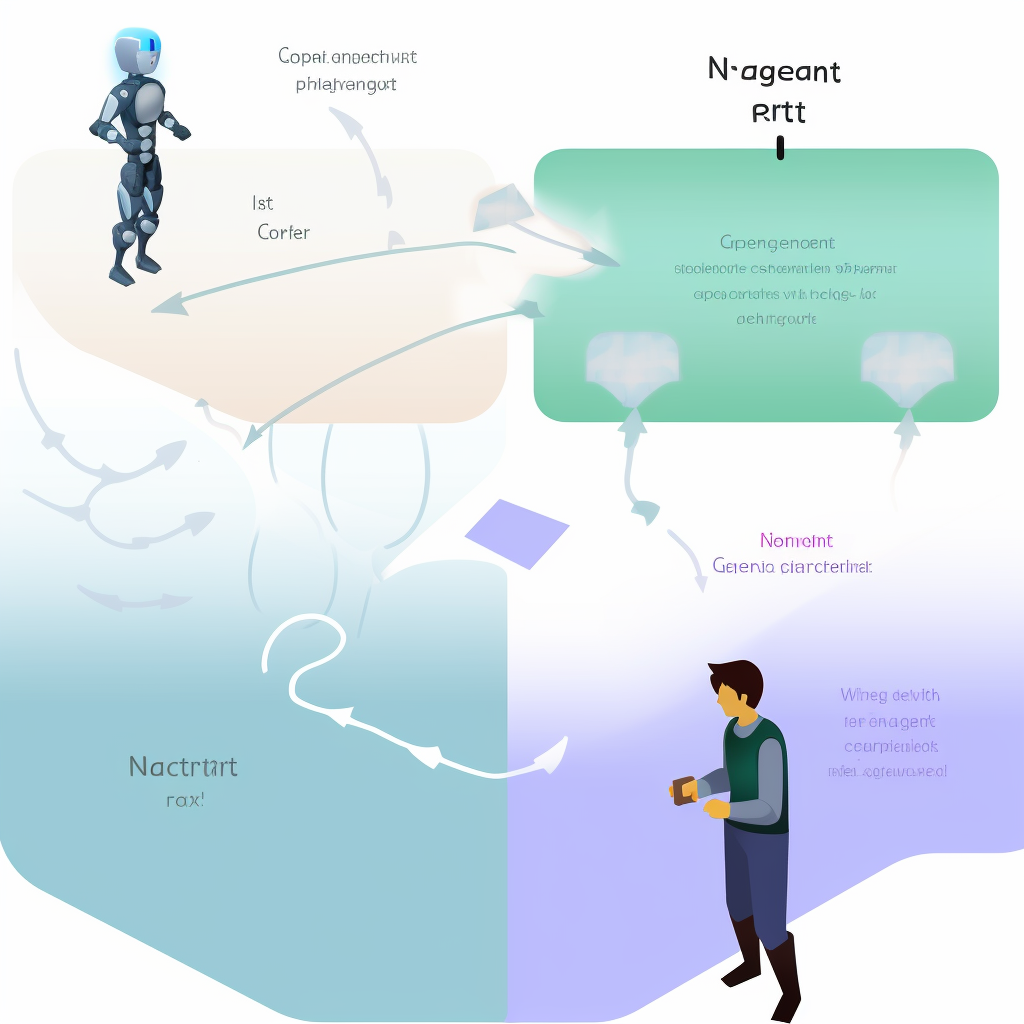}
     \caption{AI-player interaction in the virtual game.}\label{fig:game}
   \end{minipage}
\end{figure}

\subsubsection{Interactive role-playing games. (Figure~\ref{fig:game})}
GPT allows programmers to implement nuanced, interactive conversations between AI and players. These exchanges can be made to be branching and variable, giving players a one-of-a-kind and tailor-made adventure each time they log in. \textcolor{black}{Non-player characters (NPCs) backed by GPT can understand and respond to multimodal user input in continuously long interactions, providing a more immersive and customized experience}. GPT can also be used to produce random events, missions, and encounters, making the game world more dynamic and unpredictable. If used properly, GPT has the potential to greatly improve the overall quality of interactive role-playing games (NPGs).
\subsubsection{Puzzle Games.}
In the metaverse, programmers can employ GPT to make language-based \textcolor{black}{visual-based} puzzle games with interesting obstacles. The puzzle content can be generated using GPT's assistance, and it can also offer clues and solutions in a natural language setting.
\begin{figure}[!htb]
   \begin{minipage}{0.48\textwidth}
     \centering
     \includegraphics[width=.7\linewidth]{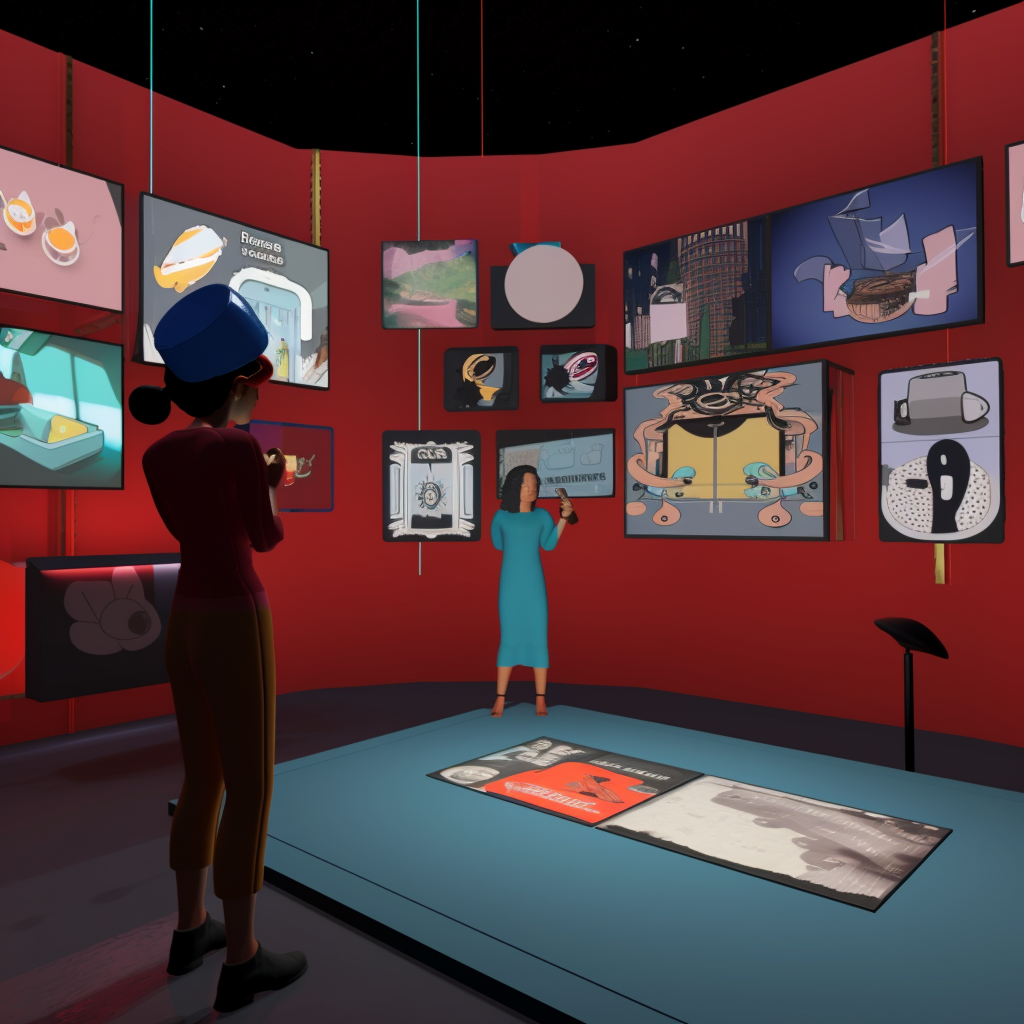}
     \caption{Interactive museum and exhibition.}\label{fig:museum}
   \end{minipage}\hfill
   \begin{minipage}{0.48\textwidth}
     \centering
     \includegraphics[width=.7\linewidth]{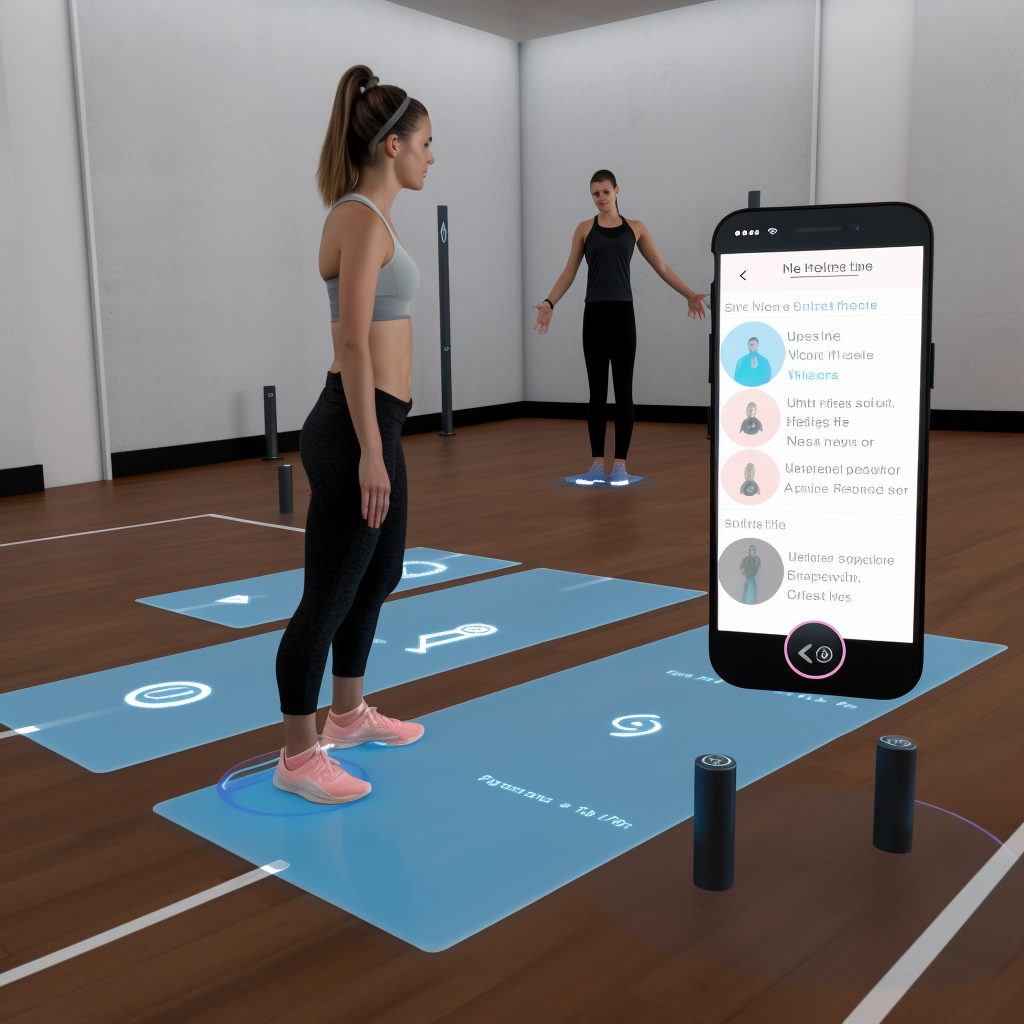}
     \caption{Personalized fitness coach}\label{fig:coach}
   \end{minipage}
\end{figure}
\subsubsection{Interactive museums and exhibitions. (Figure~\ref{fig:museum})}
GPT can build interactive museum exhibits and exhibitions in the metaverse with customized user experiences~\cite{li2019distance}. GPT may produce exhibits and information that cater to a user's particular tastes by assessing their interests and preferences, making the experience more interesting and pertinent to them.

\subsubsection{Virtual book clubs.}
GPT can customize each user's reading experience in virtual book clubs in addition to generating discussion topics and organizing virtual group discussions. GPT may produce customized book suggestions and reading lists based on a user's reading interests and history, assisting users in finding new books they are likely to enjoy reading.

\subsection{Personalization}
\subsubsection{Personalized Fitness Coaches. (Figure~\ref{fig:coach})}

GPT can be integrated into virtual fitness apps~\cite{nair2019endure} to evaluate a user's fitness development and modify their exercise regimen accordingly, offering them individualized feedback and motivation along the way. Furthermore, GPT may tailor the coaching experience by producing content that is in line with a user's unique fitness objectives, tastes, and incentives as depicted in Figure~\ref{fig:fitness}.

\begin{figure}[!htb]
   \begin{minipage}{0.48\textwidth}
     \centering
     \includegraphics[width=.7\linewidth]{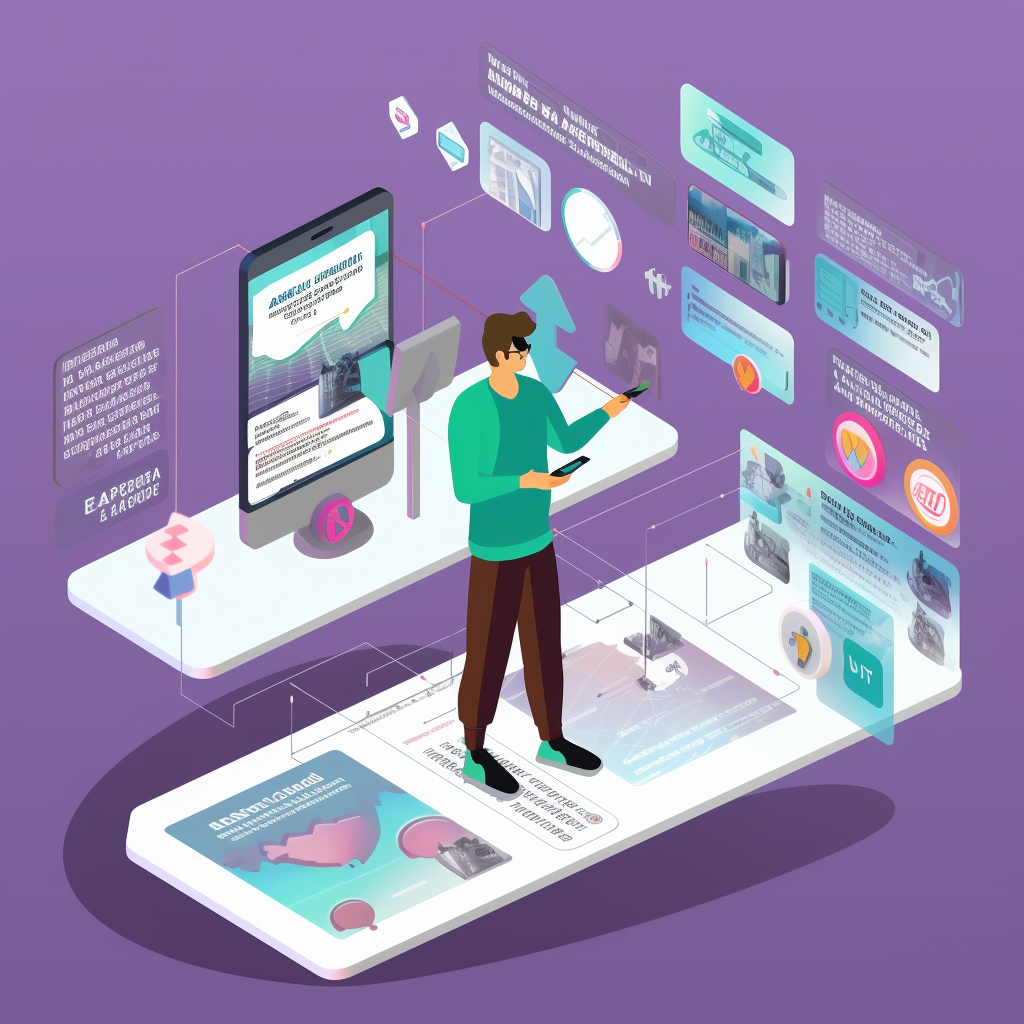}
     \caption{Personalized news and media.}\label{fig:news}
   \end{minipage}\hfill
   \begin{minipage}{0.48\textwidth}
     \centering
     \includegraphics[width=.7\linewidth]{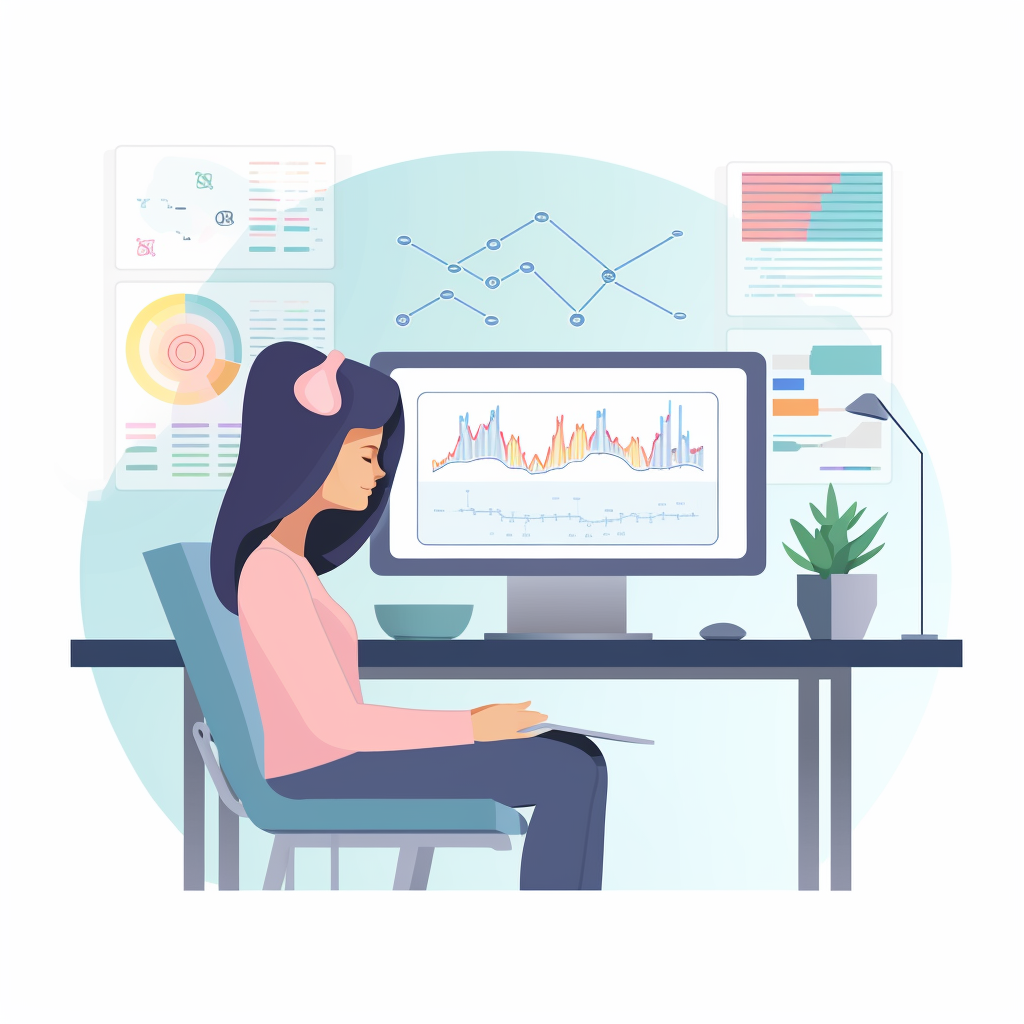}
     \caption{Virtual mental health assistant.}\label{fig:mentalhealth}
   \end{minipage}
\end{figure}

\subsubsection{Personalized news and media. (Figure~\ref{fig:news})}
Users in the metaverse can have customized news and media experiences thanks to GPT. GPT may produce customized news recommendations and summaries that are catered to each user's tastes by looking at users' interests and browsing history. Moreover, GPT may produce real-time news updates and notifications on AR glasses based on user choices and interests~\cite{lazaro2021interaction}, keeping users up to date on the most recent developments.
\subsubsection{Personalized music recommendations.}
Users in the metaverse can receive customized music recommendations through GPT. GPT may create customized playlists and recommendations that are catered to each user's tastes by examining users' listening habits and preferences. Moreover, GPT can produce explanations in natural language about why particular songs or artists are suggested, giving users a more interesting experience.

\subsection{Support}
GPT together with metaverse can also make significant contributions to providing important life support.

\subsubsection{Virtual Mental Health Assistants. (Figure~\ref{fig:mentalhealth})}

Virtual mental health aides~\cite{ma2022effectiveness} can track users' actions and emotions using data mining, and then offer tailored suggestions and interventions accordingly. Say if the assistant notices that a user is feeling stressed or concerned, it may provide relaxation techniques or advise practicing mindfulness. By producing empathy and encouraging natural language replies, together with the virtual assistant's empathetic facial expression, GPT can aid in making these assistants more sympathetic and understanding as illustrated in Figure~\ref{fig:mental}.

\subsubsection{Virtual Shopping Assistants.}
Virtual shopping assistants~\cite{billewar2022rise} offer tailored styling and fashion guidance depending on user preferences and body shape. These helpers may digitally try on clothing and accessories for consumers, offer feedback, and make suggestions via XR apps. By producing natural language responses that speak to users' emotions and preferences, GPT can assist in making these assistants more convincing and engaging. \textcolor{black}{GPT can also feedback with images to illustrate assistance and recommendations with explicit visual cues.}

\subsubsection{Virtual Personal Assistants.}
In general, virtual personal assistants can perform increasingly complex tasks and offer users in the metaverse more individualized and contextual support. These assistants, for instance, can utilize machine learning and data analysis to comprehend user preferences and behavior and then offer personalized recommendations and reminders based on that knowledge. These assistants' interactions with users could become more natural and effective if GPT is used to help make them more conversational and human-like.

\begin{figure}[!htb]
   \begin{minipage}{0.48\textwidth}
     \centering
     \includegraphics[width=.7\linewidth]{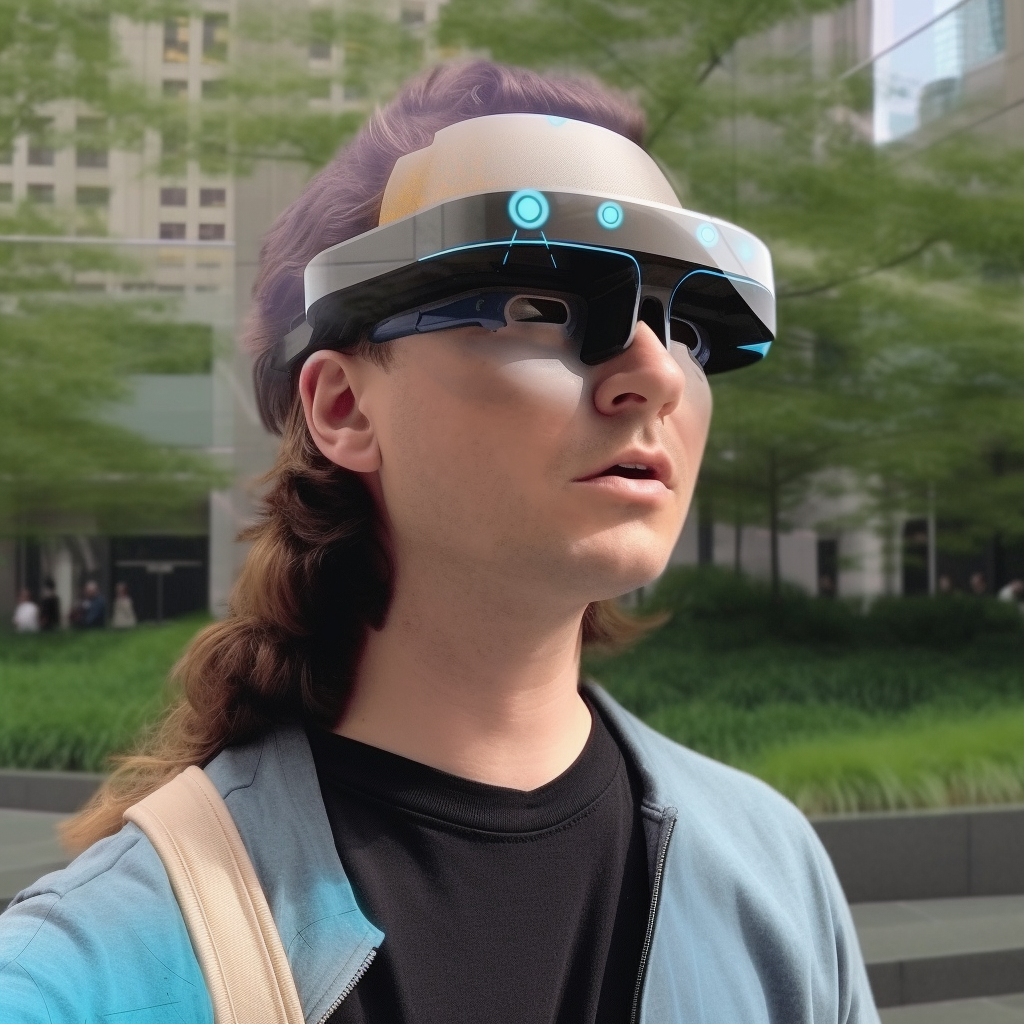}
     \caption{Virtual tour guide.}\label{fig:tour}
   \end{minipage}\hfill
   \begin{minipage}{0.48\textwidth}
     \centering
     \includegraphics[width=.7\linewidth]{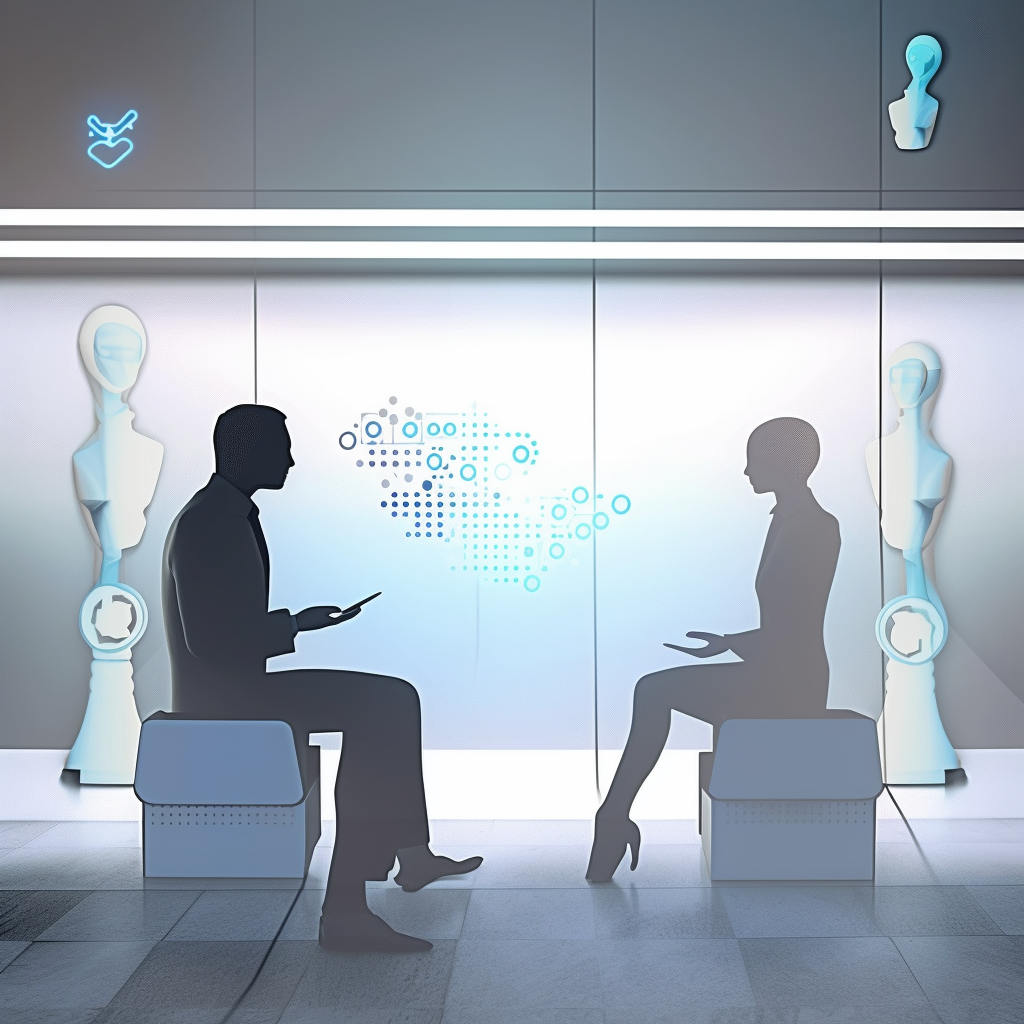}
     \caption{Language translation.}\label{fig:translation}
   \end{minipage}
\end{figure}
\subsubsection{AI-powered customer service.}
AI-powered customer service can be built using GPT to manage a variety of consumer inquiries and grievances in the metaverse. These agents can deliver effective and interesting customer service by assessing consumer inquiries and giving pertinent, natural-language responses \textcolor{black}{thanks to GPT's understanding capability and file search functionality}. For instance, if a user encounters a problem with a virtual product, an AI-powered customer support assistant can offer instant issue identification and fix, or otherwise redirection to a human representative. This can improve the customer experience while decreasing the workload of customer service staff.

\subsubsection{Virtual concierge services.}
In virtual events, GPT can be utilized to establish virtual concierge services that can offer users individualized recommendations and help. These virtual assistants can aid users with varied tasks such as scheduling, reservations, and attendance confirmations. These virtual assistants can learn more about each user's preferences over time and offer increasingly tailored recommendations.

\subsubsection{Virtual tour guides. (Figure~\ref{fig:tour})}
GPT may be used to build virtual tour guides that can give users exploring virtual sites and attractions customized recommendations and information. \textcolor{black}{These virtual assistants can comprehend complex user inquiries, such as text/audio questions and photos of the surrounding environment, thanks to the multimodal understanding and generation capability, and respond with insightful and interesting information}. Based on user interests, they can also assist consumers in finding new attractions. This can improve the user experience and motivate visitors to go further into virtual worlds.

\subsubsection{Language translation. (Figure~\ref{fig:translation})}
Real-time language translation services can be developed using GPT to enable users to converse with others in virtual worlds who speak different languages. These translation services can deliver precise and nuanced translations \textcolor{black}{to enable smooth and long conversations thanks to its long context capability}. This can increase the inclusivity and accessibility of virtual environments, facilitating global user collaboration and communication in the metaverse. Also, it can aid in removing language barriers and promote intercultural communication and comprehension.

\textcolor{black}{\subsubsection{Blockchain Assistance}
Thanks to the code interpreter and function calling functionalities, GPT can assist in smart contract writing and vulnerability identification. Leveraging the capabilities of long context, multi-agent debation, and file search, GPT can also facilitate decentralized autonomous organizations (DAOs) via writing draft proposals, summarizing discussions, and providing vote analysis for better governance.}

\section{GPT as a Destroyer}
\begin{figure}[!htb]
   \begin{minipage}{0.48\textwidth}
     \centering
     \includegraphics[width=.7\linewidth]{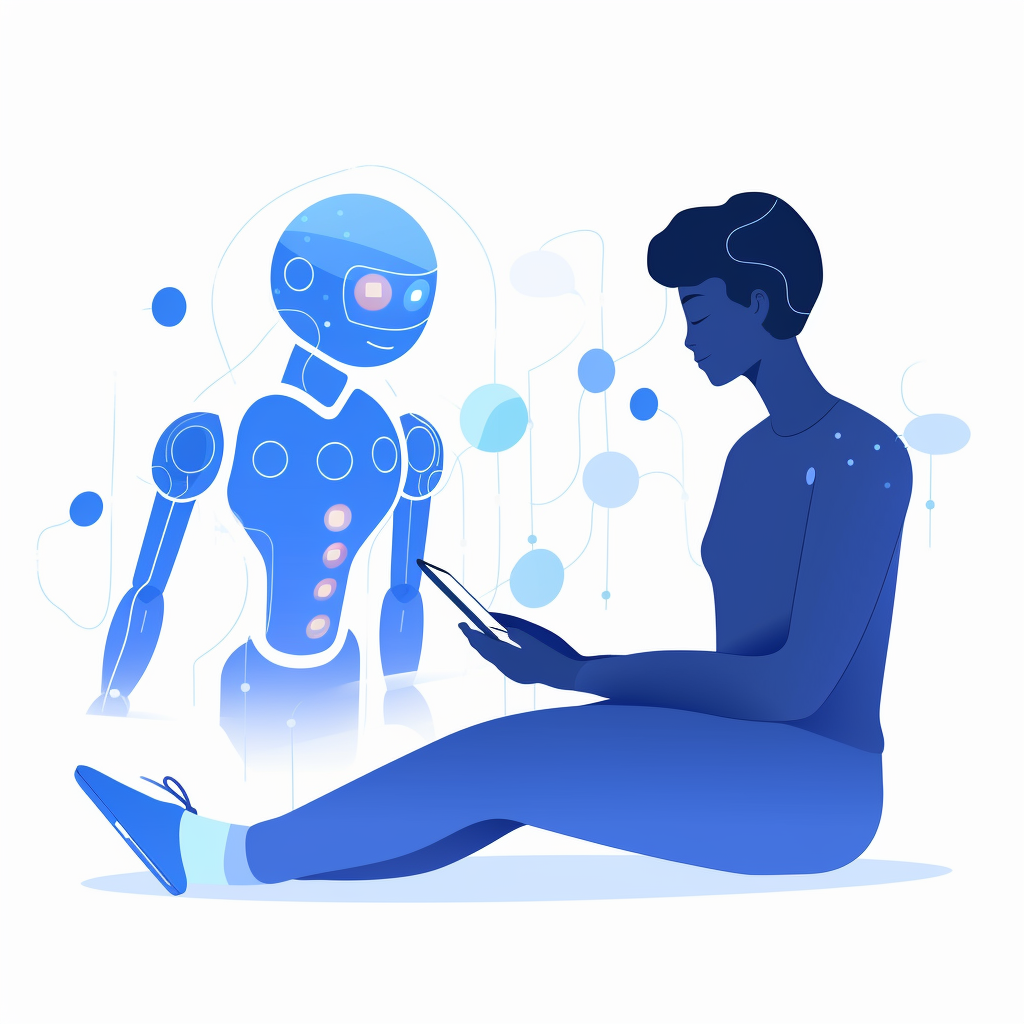}
     \caption{Less socialization.}\label{fig:detraction}
   \end{minipage}\hfill
   \begin{minipage}{0.48\textwidth}
     \centering
     \includegraphics[width=.7\linewidth]{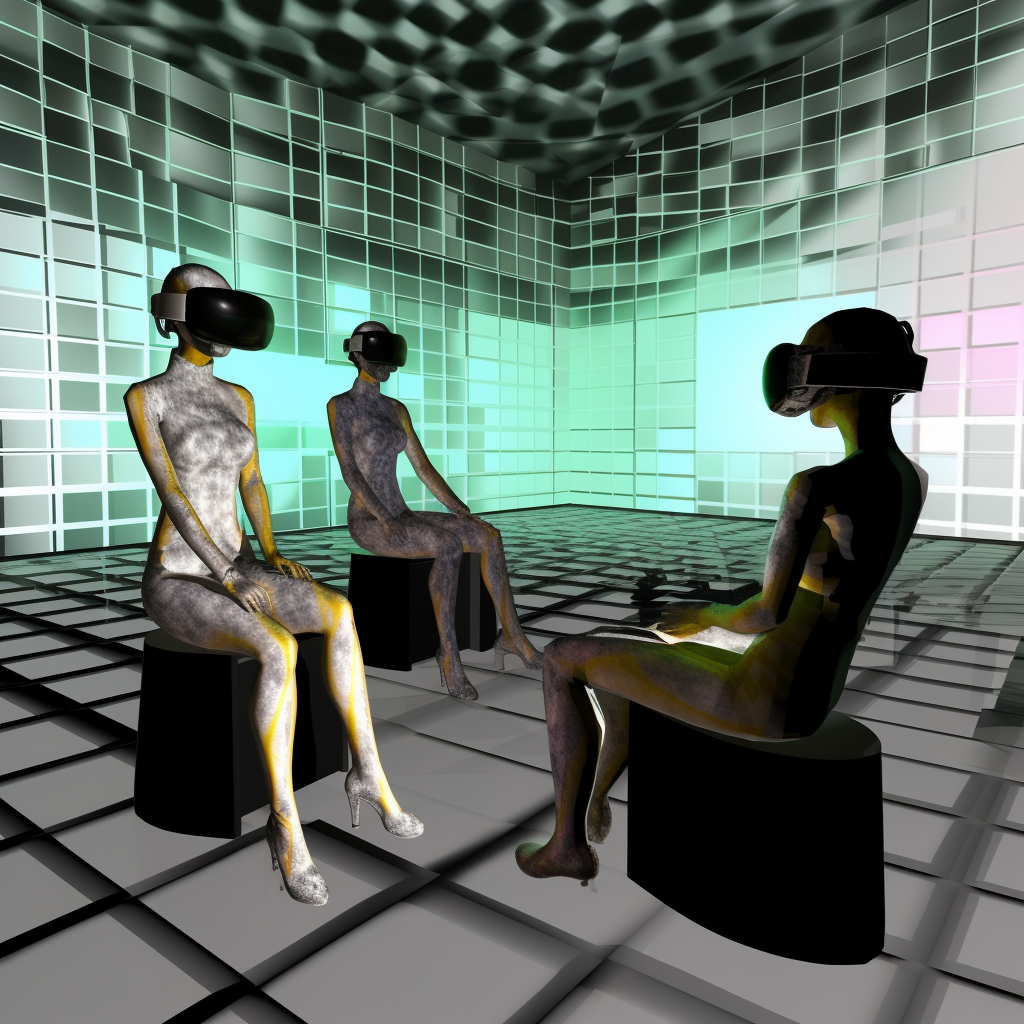}
     \caption{Virtual experience competitor.}\label{fig:competitor}
   \end{minipage}
\end{figure}

It's also possible that new technologies and services will emerge to compete with or even replace the metaverse industry. Here are a few ways that GPT or other language models could have a negative impact on the metaverse market share:

\subsection{Engagement Detraction (Figure~\ref{fig:detraction}).}
Chatbots powered by GPTs (LLMs) could potentially replace a large portion of human interaction in the metaverse, resulting in lower user engagement and socialization. Overly reliant on AI-powered interactions may impact users' engagement with the immersive experience that the metaverse is supposed to offer.
Users may become so accustomed to communicating with chatbots and miss out on human interaction nuances such as body language, facial expressions, and other nonverbal cues. This could result in a less immersive and fulfilling metaverse experience.
Furthermore, if users primarily interact with chatbots and virtual assistants, there may be less diverse interactions in the metaverse. In return, chatbots may be trained to respond in specific ways, resulting in a more limited range of interaction and less spontaneity in user-to-user communication. This could eventually undermine the immersive experience of the metaverse.

\subsection{Virtual Experience Competitor (Figure~\ref{fig:competitor}).}
GPTs (LLMs) could pave the way for creating alternative types of virtual reality experiences with lower levels of immersion than the metaverse. For example, GPT could be used to create virtual chat rooms with lower levels of 3D graphics and virtual environment design than the metaverse does. Finding such virtual chat rooms satisfying, users may be less likely to explore the more immersive virtual worlds in the metaverse. This could result in a shift from more immersive virtual worlds to simpler virtual chat rooms, which are also cheaper to build and maintain, in terms of market share.

Furthermore, if GPT can generate high-quality virtual chat rooms with minimal effort, competitors may find it easier to enter the market and create similar offerings. This could mean more competition and a smaller market share for existing metaverse companies focused on creating more complex and immersive virtual worlds.

\subsection{Formulaic Content Generation}
Contents created by GPT-like language models may become formulaic and unoriginal because, in the end, language models are trained on massive amounts of existing text which may limit their ability to create truly unique content. For example, if a virtual world within the metaverse is generated by a language model using pre-written scripts, the resulting world may feel sterile and uninspired. The virtual world's content may lack the depth and complexity that comes from human creativity and imagination, failing to engage users and keep them coming back.

\section{Moving forward}
A number of important problems, as summarized in Figure~\ref{fig:concern}, are worth addressing for the better integration of GPT\footnote{As mentioned, other LLMs are developing rapidly, so the issues broadly apply to powerful LLMs including GPT.} and the metaverse.
\begin{itemize}
    \item Privacy and data security. \textcolor{black}{GPT-enhanced metaverse can worsen the privacy and security issue compared to GPT-less metaverse (less intelligent) and GPT-enhanced internet (not immersive): 1) Powered by the long context capacity, attackers in GPT-enhanced metaverse can have longer conversations with the users or eavesdrop longer conversations between the users than in the GPT-less metaverse, thus potentially increasing the attack success rate of social engineering such as phishing and pretexting. 2) API-based fine-tuning can be leveraged by hackers to train malicious GPT models that can impersonate specific identities, e.g. CEO fraud or impersonating family members, in order to gain trust from innocent users for leaking private and sensitive data. The impersonation can include both dialogue patterns (stronger than GPT-less metaverse) and visual appearances (stronger than GPT-enhanced internet). 3) Multimodel understanding and generation capability allows the attackers to comprehend multimodal information sent by the users (stronger than GPT-less metaverse) as well as the surrounding environment (visual and audio) of the users (stronger than GPT-enhanced internet), to expose the physical location/context (AR) or virtual location/context (VR) of the users. Besides, compared to the current internet, the metaverse allows collecting more types of private data such as the avatar's embodied behavior and the user's biometric data, making the aforementioned issues more challenging.  }
    \item Defamation and fake news. GPT can generate intelligent and persuasive responses, but it is not always accurate or reliable, which means it has the potential to spread false information and fake news. Users should be taught how to critically evaluate the presented information, and researchers should take care to validate the veracity of GPT's responses.
    \item Stereotyping and negative behavior reinforcement. The language creation features of GPT have the potential to support unfavorable attitudes and prejudices. For example, programming a virtual assistant to respond to users in a passive or subservient manner can perpetuate abusive or sexist behavior. Virtual assistants and other metaverse apps must be designed in a way that promotes good habits and values. researchers must be aware of these risks.
    \item Unrealistic expectations. Users' expectations may be inflated as a result of GPT's personalized and targeted responses, particularly when it comes to education or mental health care. Users may expect virtual assistants to respond quickly and completely to complex queries or issues, but this may not always be the case. Researchers must manage user expectations and ensure that virtual assistants provide the appropriate levels of assistance.
    \item Dependence on technology. Using GPT and other virtual assistants in the metaverse may lead to a greater reliance on technology and less on interpersonal contact. As a result, socialization, communication, and emotional health may all suffer. Researchers must consider these risks when creating metaverse apps, balancing the benefits of technology with the importance of human connection and interaction.
\end{itemize}
\textcolor{black}{Relevant progresses:
\begin{itemize}
    \item \textbf{Metaverse perspective.} EU has launched several initiatives for metaverse regulations\footnote{\url{https://www.insideprivacy.com/metaverse/regulating-the-metaverse-in-europe/}}. As stated in the initiative in May 2023\footnote{\url{https://digital-strategy.ec.europa.eu/en/library/eu-initiative-virtual-worlds-head-start-next-technological-transition}}, data privacy protection is one of the major concerns and is continuously expected to be guaranteed by General Data Protection Regulation (GDPR). Similarly, other existing laws and regulations such as the California Consumer Privacy Act (CCPA) in the United States can also keep serving data privacy protection in the metaverse. However, as the World Economic Forum's ``Defining and Building the Metaverse'' initiative noted, the metaverse will probably need a broader privacy strategy to deal with the richer data collection. So far, there seem to be some actions being taken to make regulations and policies tailored to the metaverse, but none have been announced. \textit{This is partly due to the practical challenges of developing new regulations and policies for a realm that is still developing, also due to the reactive nature of regulation hence further regulation will likely be proposed only when the metaverse is more deployed.}
    \item \textbf{AI perspective.} OpenAI has gone through some dramatic changes\footnote{\url{https://www.nature.com/articles/d41586-023-03700-4}} about ethical AI surrounding aforementioned challenges such as privacy and data security\footnote{\url{https://www.reuters.com/technology/cybersecurity/italy-regulator-notifies-openai-privacy-breaches-chatgpt-2024-01-29/}} and fake news\footnote{\url{https://www.nature.com/articles/d41586-023-02990-y}}. Relevant concerns are raised especially in the EU, e.g., the Italian Data Protection Authority temporarily blocked ChatGPT in March 2023 due to its violation of GDPR. Although the ban was lifted after OpenAI implemented the required changes, many countries have followed Italy to do similar investigations on ChatGPT\footnote{\url{https://www.edpb.europa.eu/system/files/2024-05/edpb_20240523_report_chatgpt_taskforce_en.pdf}}. Currently, OpenAI claims their products comply with privacy laws such as GDPR and CCPA, and offer a ``Data Processing Addendum'' for governing the processing of user data. Specific actions include a ``Privacy Request Portal'' which allows the users to opt out of having their data used to improve OpenAI's non-API services\footnote{\url{https://privacy.openai.com/policies}}. Users can also disable chat history and model training\footnote{\url{https://help.openai.com/en/articles/8983082}}.
\end{itemize}}

\textcolor{black}{
\textbf{Suggestions.} As metaverse and GPT (or LLMs in general) are both still in their early stages, reactive development of regulations and policies would most likely experience a delay. A more practical method, as many are employing, is to reuse existing regulations and policies such as GDPR and CCPA and check if refinement or expansion is needed for specific products or use cases. In general, regulations on AI are more standardizable compared to the metaverse, because most AI-related services today are using data from the same area of the services provided to cater to local markets and tastes, while the metaverse (the ultimate version) is a decentralized, global scale realm consisting of countless XR platforms running by different companies from different countries. Therefore, one-regulation-for-all would likely fail to tackle the aforementioned issues in the cross-area/country metaverse, e.g., different countries have different privacy requirements reflecting on different data collection and usage restrictions on users from different countries but ``live'' in the same metaverse. \textit{While GPT (LLMs) can make it worse by providing more means to use the data (for good and for bad purposes), it can also become an effective helper to deploy per-user regulation assistance that warns and suggests users with regulations or actions tailored to different countries' laws and users' preferences. API-based fine-tuning allows metaverse service providers or even users themselves to train ``regulation assistance'' or ``watchdog'' to protect users from aforementioned issues.}
}

\section{Conclusion}
The metaverse industry is still in its infancy, it's unclear what kinds of virtual experiences will be most popular and profitable in the long run at this point. GPT-powered services could potentially complement rather than compete with metaverse services and could help to expand the overall market for metaverse experiences. Furthermore, GPT can be used to supplement and enhance human creativity in the metaverse. As the metaverse develops, so will opportunities for GPT and other forms of NLP in MMORPGs. The need for interactive and unique experiences will grow as more people join the metaverse. GPT is just one example of a natural language processing and generation technology that may be used to build a variety of useful and entertaining virtual assistants and companion applications. GPT's potential for facilitating the development of sophisticated conversational agents in the metaverse is an intriguing avenue for potential future study. With reinforcement learning and other methods, these agents could acquire a deeper comprehension of user actions and reactions to provide more tailored and interesting interactions. 
While researchers have an amazing chance to create individualized and engaging virtual experiences by combining GPT with the metaverse, they need to be aware of the obstacles to make the virtual world a safe and ethical place for all users.
\bibliography{iclr2022_conference}
\bibliographystyle{iclr2022_conference}
\appendix
\begin{figure*}
    \centering
    \includegraphics[width=\textwidth]{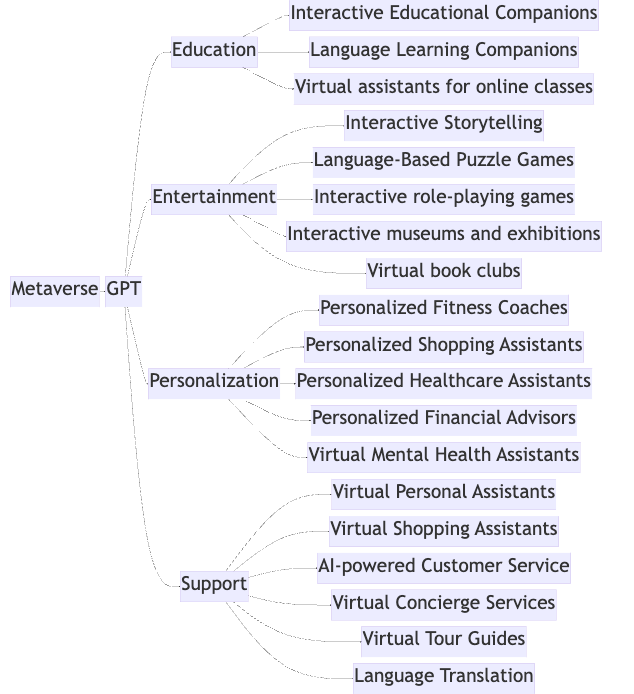}
    \caption{GPT as a ``Savior'' for metaverse}
    \label{fig:savior}
\end{figure*}
\begin{figure*}
    \centering
    \includegraphics[angle=90,origin=c,scale=0.8]{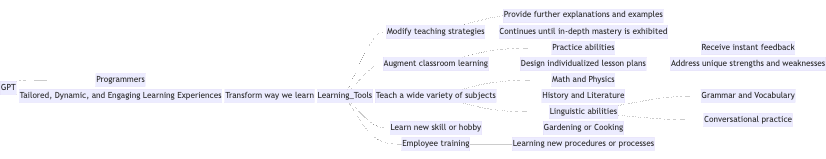}
    \caption{Interactive educational companions.}
    \label{fig:companion}
\end{figure*}
 \begin{figure}
    \centering
    \includegraphics[width=.8\columnwidth]{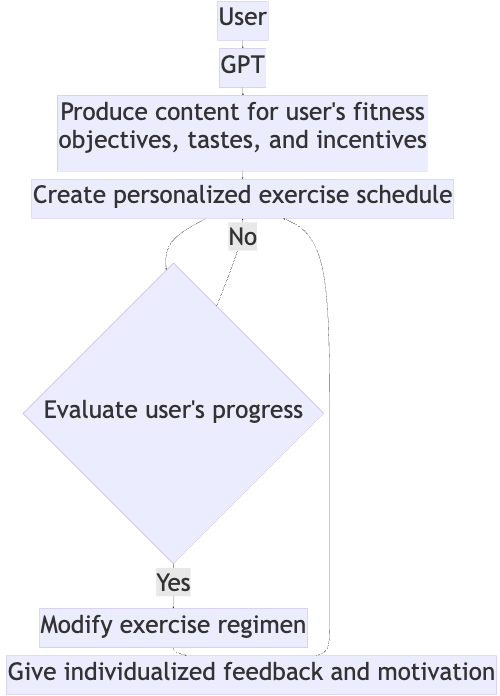}
    \caption{Personalized fitness coaches.}
    \label{fig:fitness}
\end{figure}
\begin{figure}
    \centering
    \includegraphics[width=.8\columnwidth]{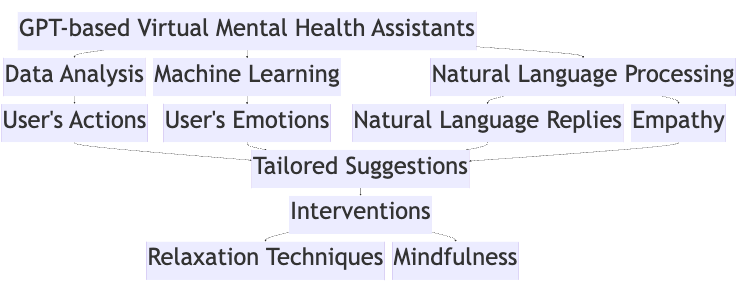}
    \caption{Virtual mental health assistants.}
    \label{fig:mental}
\end{figure}

\begin{figure*}
    \centering
    \includegraphics[angle=90,origin=c,scale=0.65]{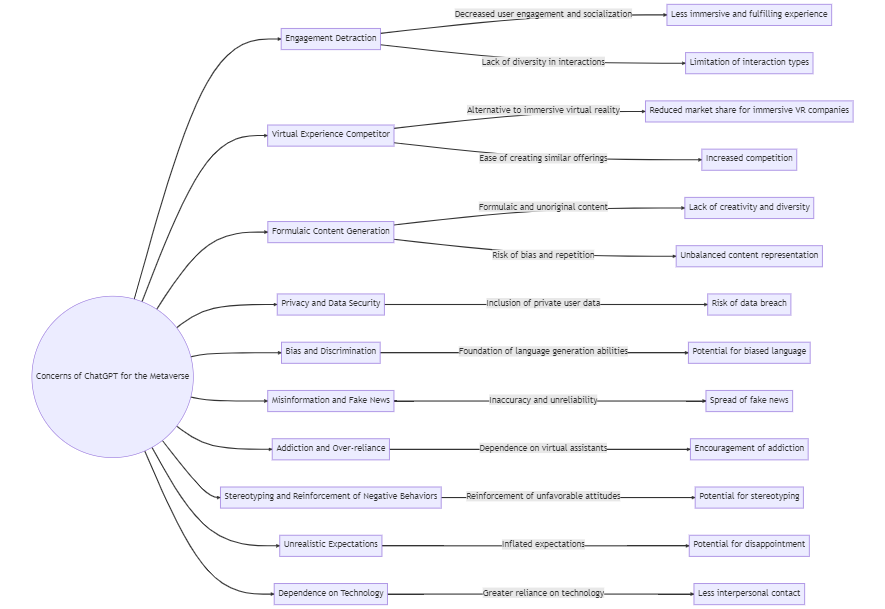}
    \caption{GPT as a ``Destroyer'' for the metaverse.}
    \label{fig:concern}
\end{figure*}
\end{document}